\documentclass[12pt,a4paper]{article}
\pdfoutput=1

\usepackage{physics}
\usepackage{amssymb}
\usepackage{amsmath}
\usepackage{amsfonts}
\usepackage{amsthm}
\usepackage{setspace}
\usepackage{color}
\usepackage[pdftex]{graphicx}
\usepackage{authblk}
\usepackage{subfigure}
\usepackage{float}
\usepackage[utf8]{inputenc}

\usepackage{datetime}

\numberwithin{equation}{section}

\theoremstyle{plain}

\usepackage{color}

\newdateformat{daymonthyear}{ \monthname[\THEMONTH] \THEDAY, \THEYEAR}
\newdateformat{monthyear}{\monthname[\THEMONTH] \THEYEAR}

\begin{document}

\title{Meissner State in Classical Ideal Charged Fluid}

\author{Vladimir Toussaint$^{1}$\thanks{{\tt vladimir.toussaint@nottingham.edu.cn}} }
\author{George H. Goedecke$^{2}$\thanks{\tt ggoedeck@nmsu.edu}}
\affil{$^{1}$School of Mathematical Sciences, University of Nottingham Ningbo China,\\
 Ningbo 315100, China}
\affil{$^{2}$Physics Department, New Mexico State University, Las Cruces, NM, USA}
\date{\daymonthyear\today}

\maketitle

\begin{abstract}
It is shown that a model of a superconductor as a classical ideal charged fluid at constant uniform temperature in a uniform compensating positively charged background exhibits the Meissner state with the correct value of the penetration depth. The differences between this model and the textbook classical model that fails to predict the Meissner state are discussed, as are the similarities of this model to the coupled Schr\"{o}dinger-Maxwell model, which does predict flux exclusion. In particular, this model does not neglect the pressure term and terms that are bilinear in the velocity in the fluid equations; these terms are essential to the results.
\end{abstract}


\newpage

\section{Introduction\label{sec:intro}}

It is a widely held belief that the Meissner state, the exclusion of magnetic flux by a (type I) superconductor, cannot be derived from a classical treatment of collissionless charge carriers, but instead requires a separate ansatz, such as that made by F. and H. London  \cite{London:F&H}  or by Ginsburg and Landau \cite{Ginzburg:1950sr}. Here, we show that the Meissner state actually results directly from a model of a superconductor as a classical ideal charged fluid at constant uniform temperature in a uniform compensating positively charged background.
%

In what follows, in Sec. II we develop the model, in Sec. III we discuss its gauge transformation properties, in Secs. IV and V we contrast the model with the classical textbook model and the Schr\"{o}dinger-Maxwell model, and in Sec. VI we offer a brief conclusion and discussion.

\section{Classical Ideal Charged Fluid Model\label{sec:UDW-detector}}

The fluid equations for this model are (cgs units)

\begin{subequations}
\begin{align}
\label{FluidEqnA}
m \left( \frac{\partial \pmb{v}}{\partial t} + \pmb{v}\cdot\nabla \pmb{v}      \right) &= -n^{-1}\nabla p + q (\pmb{E} + c^{-1}\pmb{v}\times \pmb{B}) \, , \\
\frac{\partial n}{\partial t} + \nabla \cdot (n\pmb{v}) &= 0\, ,
\label{FluidEqnB}
\end{align}
\end{subequations}
where $(n, \textbf{v})$ are the fluid (number density, velocity) fields, $(q,m)$ are the (charge, mass) of the fluid particles, $p$ is the fluid pressure, and $(\pmb{E}, \pmb{B)}$ are the self-consistent (electric, magnetic) fields that satisfy the Maxwell equations (Gaussian units)
\begin{subequations}\label{SelfConsitent:MaxwellFields}
\begin{align}
\label{Divergence:E&Bfields}
\nabla \cdot \pmb{E} &= 4 \pi q (n - n_0) \, ,   &   \nabla\cdot \pmb{B} &= 0\, ,      \\
\nabla \times \pmb{E} &= -c^{-1} \frac{\partial \pmb{B}}{\partial t} \, ,   &  \nabla \times \pmb{B} &= c^{-1} \left( \frac{\partial \pmb{E}}{\partial t}  + 4 \pi q n\pmb{v}\right)\, .
\label{Curl:E&Bfields}
\end{align}
\end{subequations}
Here, $n_0$ is the space-time average of the number density $n(\pmb{r},t)$. Note that in \eqref{FluidEqnA}, the usual phenomenological collision term, $(-m\nu\pmb{v})$, is omitted from the right-hand side, in accord with this fluid model in which the collision frequency $\nu = 0$.   

We write the electromagnetic fields in terms of their potentials in the usual way,
\begin{align}\label{Emags:Fields}
\pmb{E} = -c^{-1}\frac{\partial \pmb{A}}{\partial t} -\nabla \phi, \quad \quad    \pmb{B} = \nabla \times \pmb{A}\, ,
\end{align}
and also define the ``canonical momentum field " $\pmb{P}$ by 
\begin{align}\label{CanonicalMomentField}
\pmb{P} = m \pmb{v} + q c^{-1}\pmb{A}\, .
\end{align}
Substituting these in \eqref{FluidEqnA}, we get exactly
\begin{align}\label{MomentumFieldEq:Fluid}
\frac{\partial{\pmb{P}}}{\partial t} -\pmb{v}\times(\nabla\times \pmb{P}) = -n^{-1}\nabla p - \nabla\left( \frac{1}{2} m v^2 + q \phi   \right) \, .
\end{align}

We recall that, for uniform constant temperature, we may write
\begin{align}\label{UniformConst:ChemiPo}
n^{-1}\nabla{p} = \nabla\mu(n) \, ,
\end{align}
where $\mu(n)$ is the chemical potential; i.e., $n^{-1}\nabla p$ is a gradient. Using this in \eqref{MomentumFieldEq:Fluid}, we see that the canonical momentum field $\pmb{P}$ may always be chosen to be a gradient. Thus we get from \eqref{Emags:Fields} and \eqref{CanonicalMomentField}
\begin{align}\label{velocityField:MagField}
\nabla\times\pmb{v} = -\frac{q\pmb{B}}{mc}\, ,
\end{align}
and from \eqref{MomentumFieldEq:Fluid} and \eqref{UniformConst:ChemiPo}
\begin{align}\label{MomentumFieldEq:FluidB}
\frac{\partial{\pmb{P}}}{\partial t} = -\nabla \left(\mu(n) + \frac{1}{2} m v^2 + q \phi   \right) \, .
\end{align}

We expect that $|n- n_0| \ll n_0$ for most situations\footnote{This might not be true if $n_0\rightarrow 0$, as should happen near a superconducting-normal transition.}, because charge separations in a charged fluid are very difficult to manintain, and also because \eqref{Divergence:E&Bfields} and \eqref{MomentumFieldEq:FluidB} show that $|n- n_0| \propto v^2$, which should be small. Therefore, for such situations, it should be a very good approximation to replace $n$ by $n_0$ in \eqref{Curl:E&Bfields}, whereby, combining \eqref{SelfConsitent:MaxwellFields} and \eqref{velocityField:MagField}, we get immediately
\begin{subequations}\label{WaveEq:FieldFluid}
\begin{align}
\Box\pmb{B}  - \lambda_p^{-2}\pmb{B} = 0\, ,
\end{align}
where
\begin{align}\label{Penetration:Length}
 \lambda_p  \equiv \left( \frac{mc^2}{4\pi n_0 q^2} \right)^{\frac{1}{2}} \,
\end{align}
\end{subequations}
is the penetration length. The static limit of \eqref{WaveEq:FieldFluid} is the same as the result of the London theory.

\section{Gauge Properties of the Classical Model}

Since $\pmb{P}$ is a gradient, we may write \eqref{CanonicalMomentField} as
\begin{align}
\pmb{P} = \nabla J = m \pmb{v} + q c^{-1}\pmb{A} \, ,
\end{align}
where $J$ has the dimension of angular momentum. This result suggests that we make a gauge transformation in an attempt to eliminate $J$ if it should happen to be non-zero: We let
\begin{align}\label{GaugeTransfrom:Vector&ScalarPotentials}
\pmb{A} = \pmb{A}' + cq^{-1}\nabla J \, , \quad \quad \phi = \phi' - q^{-1}\frac{\partial J}{\partial t} \, ,
\end{align}
and for convenience require the Lorentz gauge for the primed potentials, which yields
\begin{align}
\nabla\cdot \pmb{A} + c^{-1}\frac{\partial \phi}{\partial t} - cq^{-1}\Box J = 0 \, .
\end{align}
The Maxwell equations \eqref{SelfConsitent:MaxwellFields} and \eqref{FluidEqnB},  \eqref{Emags:Fields}, \eqref{CanonicalMomentField}, \eqref{MomentumFieldEq:FluidB} in this gauge then reduce exactly to 
\begin{subequations}\label{GaugeTransformedEqs}
\begin{align}
\label{WaveEq:VectorPot}
\Box \pmb{A}' - \frac{4\pi q^2n}{mc^2}\pmb{A}' &= 0 \, , \\
\label{WaveEq:ScalarPot}
\Box \phi'  - 4\pi q(n-n_0) &= 0 \, , \\
\label{ChemicalPotCoupledPhiField}
\nabla\left(\mu(n) + \frac{1}{2}mv^2 + q\phi'      \right) &= 0 \, , \\
\label{Rcoupledtovfield}
\frac{\partial}{\partial t}\left( R - \frac{q\phi'}{mc^2}     \right) + (\nabla R)\cdot \pmb{v} &= 0 , \quad  R \equiv \ln \left(\frac{n}{n_0}\right) \, , \\
\label{velocityField:GaugeForm}
\pmb{v} &= -\frac{q\pmb{A}'}{mc} \, , \\
\label{LorentzGauge}
\nabla \cdot\pmb{A}' + c^{-1}\frac{\partial \phi'}{\partial t} &= 0\, , \\
\pmb{B} = \nabla \times \pmb{A}' \, ,\quad \quad   \pmb{E} &= -c^{-1}\frac{\partial \pmb{A}'}{\partial t} -\nabla \phi'   \, .
\label{E&BintermsofScal&vectorPoten}
\end{align}
\end{subequations}
Note that not only does $J$ disappear from \eqref{E&BintermsofScal&vectorPoten}, but also from the fluid motion equation  \eqref{MomentumFieldEq:FluidB}, which converts to \eqref{ChemicalPotCoupledPhiField}. So we may choose this gauge in simply connected regions. Note also that, in the static limit, these equations imply that $n\neq n_0$ and $\phi' \neq$ constant, but that $\pmb{A}', \pmb{v}$ and $\pmb{B}$ still go to zero inside the fluid, within a short distance from the surface, unless $n_0\rightarrow 0$.

However, in general by Stokes' theorem we must have
\begin{align}\label{MagneticFlux}
\int_C\pmb{A}\cdot d\pmb{l} = \Phi \, ,
\end{align}
where $\Phi$ is the total magnetic flux enclosed by the arbitrary closed path $C$. Consider a non-simply connected region of the fluid. In particular, imagine a cylindrical annulus of ideal fluid whose thickness is much greater than $\lambda_p$, and let there be a magnetic field directed along the axis inside the inner cylinder (vacuum), so that $\Phi =0$ for a closed path in this fluid. For example, this is the situation for an infinitely long current-carrying circular solenoid in vacuum, surrounded by a superconductor. If we solve the static limit of \eqref{WaveEq:VectorPot} to \eqref{E&BintermsofScal&vectorPoten} for such a case, we find that both $\pmb{v}$ and $\pmb{A}'$ go to zero deep within the fluid. Therefore this $\pmb{A}'$ is inadequate: Although it yields the correct $\pmb{B}$ inside the fluid, it cannot satisfy \eqref{MagneticFlux}. We must add to it a term $cq^{-1}\nabla J$, as in \eqref{GaugeTransfrom:Vector&ScalarPotentials}, where $J = (q\Phi/2\pi c)\theta$, and $\theta$ is the azimuthal angle in cylindrical coordinates. This will satisfy \eqref{MagneticFlux} without influencing the results of  \eqref{GaugeTransformedEqs}. Thus, we see that, even classically, $J$ cannot be transformed away in such situations.\footnote{In quantum mechanics, we know that $J$ is proportional to the phase of the superconducting wavefunction, or $J = l\hbar \theta$, where $ l $ is an integer and $\theta$  is the azimuthal angle in cylindrical coordinates.  This leads to quantization of the enclosed flux, $\Phi = (hc/q) l$. Also we know that this behaviour is the origin of the Aharonov-Bohm effect. For a fuller discussion, see \cite{Ingram}.}  

\section{Comparison with Classical Model}
The model developed above is fundamentally different from the classical models discussed in many texts \cite{Gerald Burns, Neil&David, Phlilp, Harald}, which treat the mobile charged particles as noninteracting and having uniform number density, and which also omit the $\pmb{v}\times\pmb{B}$ term in the Lorentz force and of course omit the $\pmb{v}\cdot \nabla\pmb{v}$ term in the convective derivative, since they don't use a fluid treatment. As we have seen, the collective behavior implied by the fluid equations is essential to our results; the $\pmb{v}\times\pmb{B}$,  $\pmb{v}\cdot \nabla \pmb{v}$, and pressure term are all of central importance. In fact, if one regards these terms and the pressure term, which are $O(v^2)$, as negligibly small, and drops them \textit{a priori}, then \eqref{FluidEqnA} reduces to $\partial \pmb{v}/\partial t = \pmb{E}/m$. Then, from the Maxwell equations \eqref{Curl:E&Bfields}  one gets only the time derivative of \eqref{velocityField:MagField} and then only the equation $(\Box - \lambda_p^{-2})\partial\pmb{B}/\partial t = 0$, satisfied by any $\pmb{B}$ that is independent of time. This is the textbook dilemma.

\section{Comparison with Quantum Model}
Perhaps the simplest quantum mechanical model is that given by Feymann \cite{Feymann}. He starts with the Schr\"{o}dinger equation
\begin{align}
i\hbar\frac{\partial \psi}{\partial t} =\frac{1}{2m}\left( -i\hbar\nabla-qc^{-1}\pmb{A}   \right)^2\psi   + q \phi\psi \, ,
\end{align}
and assumes that the particles being described are Cooper pairs that behave as bosons. Thus, all of them may be in the same (ground) state, whereby the (charge, current) densities $(\rho, \pmb{j})$ are given by
\begin{align}
\rho = q |\psi|^2, \quad \pmb{j} =\frac{q}{2m} \psi^*(-i\hbar\nabla - qc^{-1}\pmb{A})\psi +\text{c.c.}
\end{align}
He then shows that the coupled Schr\"{o}dinger-Maxwell equations are equivalent to the ideal charged fluid equations which are strikingly similar to \eqref{FluidEqnA} and \eqref{Divergence:E&Bfields}. In fact, if in our motion equation \eqref{MomentumFieldEq:FluidB} we replace the chemical potential $\mu(n)$ by $(\hbar^2/2m)(\nabla^2 n^{\frac{1}{2}})/n^{\frac{1}{2}}$, then our equations are equivalent to the hydrodynamical form of the Sch\"{o}dinger-Maxwell equations, from which the Meissner effect does result.\footnote{Complete equivalence of the solutions of our classical equations with those of  Schr\"{o}dinger-Maxwell equations would require more, namely, that the function $\psi \equiv n^\frac{1}{2}\exp(iJ/\hbar)$ be single-valued and satisfy the usual boundary conditions. In the  Schr\"{o}dinger-Maxwell treatment, the canonical momentum field is a priori a gradient.} Curiously, the derivation of the Schr\"odinger equation from classical statistics \cite{G.George} yields exactly Feynman's equation for a collection of charged bosons all in the ground state.

\section{Discussion and Conclusion}
We have shown that a completely classical treatment of an ideal charged fluid does predict exclusion of static magnetic fields. The crucial element of the proof lies in showing that the canonical momentum field $\pmb{P}$ defined in \eqref{CanonicalMomentField} is a gradient, even in a classical treatment. 

If a conventional collision term $(-m\nu\pmb{v})$ with collision frequency $\nu >0$ is included on the right-hand side of Eq. \eqref{FluidEqnA}, then $\pmb{P}$ cannot be a gradient or be identically zero (see Appendix~\ref{Collision}), and  then $\pmb{v} = 0$ rather than $\pmb{P} =0$ is the classical static limit result in simply connected regions. 

Note that our approximate expression \eqref{WaveEq:FieldFluid} and the exact epression \eqref{WaveEq:VectorPot} imply that, for frequencies greater than the plasma frequency $\omega_p \equiv \lambda_p/c$, electromagnetic waves can propagate in the fluid, just as in collisionless plasma.

However; our model has several  limitations. For example, it does not account for the full Meissner effect: the exclusion of the magnetic flux during the superconducting phase transition. During such a dynamical process the system is not at a uniform constant temperature, whereby \eqref{UniformConst:ChemiPo} would not be applicable. In this case, the canonical momentum field $\pmb{P}$ could not be chosen as a gradient field. Nonetheless, 
this classical model suitably describes the Meissner state wherein a superconductor has no internal magnetic field.

Finally, we note that neither this nor any other classical treatment can offer an explanation of the absence of electron-lattice interaction in the superconducting state; for that, one needs the insight and power of quantum mechanics and the BCS theory concept of Cooper pairs \cite{Bardeen:1957mv} or some other mechanism that provides an energy gap between the electronic ground state and excited states. 

\section*{Acknowledgments}

We thank James M. Wilkes and Richard L. Ingraham  for helpful discussions.

\appendix

\section{Inclusion of Collision Term\label{Collision}}
If we decide to include the collision term $-m\nu\pmb{v}$ on the right hand side of \eqref{FluidEqnA} then we obtain instead of \eqref{MomentumFieldEq:Fluid},
\begin{align}\label{ModifiedMomentumFieldEq:Fluid}
\frac{\partial{\pmb{P}}}{\partial t} -\pmb{v}\times(\nabla\times \pmb{P}) = -n^{-1}\nabla p - \nabla\left( \frac{1}{2} m v^2 + q \phi   \right) -m\nu\pmb{v}\, .
\end{align}
For a uniform constant temperature, $n^{-1}\nabla{p}$ is a gradient field as shown in \eqref{UniformConst:ChemiPo}. When $\nu =0$, the canonical momentum field $\pmb{P}$ can be chosen to be a gradient field without loss in generality.   
If the collision term is nonzero but small, say $|m\nu\pmb{v}| \ll 1$, suppose then that the canonical momentum is a gradient field plus a small correction term $\pmb{\epsilon}(\pmb{r},t)$, where $|\pmb{\epsilon}(\pmb{r},t)|\ll 1$, then we obtain instead of \eqref{velocityField:MagField} and \eqref{MomentumFieldEq:FluidB}
\begin{align}\label{Modified_velocityField:MagField}
\nabla\times\pmb{v} &= -\frac{q\pmb{B}}{mc} + m^{-1}\nabla\times \pmb{\epsilon} \, , \\
\frac{\partial{\pmb{P}}}{\partial t} &= \pmb{v}\times(\nabla\times \pmb{\epsilon}) -\nabla \left(\mu(n) + \frac{1}{2} m v^2 + q \phi   \right) -m\nu\pmb{v} \, .
\end{align}
 In the approximation where  $|n- n_0| \propto v^2 \ll 1$, whereupon $n$ can be replaced with $n_0$ in \eqref{Curl:E&Bfields}, we obtain by collecting \eqref{SelfConsitent:MaxwellFields} and \eqref{Modified_velocityField:MagField}, 
\begin{align}\label{ModifiedWaveEq:FieldFluid} 
\Box\pmb{B}  - \lambda_p^{-2} \left( \pmb{B} -cq^{-1}\nabla\times \pmb{\epsilon} \right) = 0\, ,
\end{align}
where $\lambda_p$ is the penetration length given in \eqref{Penetration:Length}. Therefore, only in the limit where the collision term vanishes identically do we actually recover the  result of London theory in the static limit of \eqref{ModifiedWaveEq:FieldFluid}.

\end{document}